\documentstyle[12pt]{article}                    
\newfont{\ffont}{msym10}                        
\newcommand{\beq}{\begin{equation}}             
\newcommand{\eeq}{\end{equation}}               
\newcommand{\bqry}{\begin{eqnarray}}            
\newcommand{\eqry}{\end{eqnarray}}              
\newcommand{\bqryn}{\begin{eqnarray*}}          
\newcommand{\eqryn}{\end{eqnarray*}}            
\newcommand{\NL}{\nonumber \\}                  
\newcommand{\preprint}[1]{\begin{table}[t]      
            \begin{flushright}                  
            \begin{large}{#1}\end{large}        
            \end{flushright}                    
            \end{table}}                        
\newcommand{\PD}[2]                             
    {\frac{\partial^{#2}}{\partial #1^{#2}}}    
\begin{document}
\preprint{LA-UR-97-XXXX}
\title{Towards Resolution of the \\ Scalar Meson Nonet Enigma}
\author{\\ L. Burakovsky\thanks{E-mail: BURAKOV@PION.LANL.GOV} \
\\  \\  Theoretical Division, T-8 \\  Los Alamos National Laboratory \\ Los
Alamos NM 87545, USA \\  \\  \\  \\ }
\date{{\it Dedicated to Prof. L.P. Horwitz \\ 
on the occasion of his 65th anniversary}}
\maketitle
\begin{abstract}
By the application of a linear mass spectrum to a composite system of both the
pseudoscalar and scalar meson nonets, we find three relations for the masses 
of the scalar states which suggest the $q\bar{q}$ assignment for the scalar 
meson nonet: $a_0(1320),$ $K_0^\ast (1430),$ $f_0(1500)$ and $f_0^{'}(980).$ 
\end{abstract}
\bigskip
{\it Key words:} hadronic resonance spectrum, quark model, scalar mesons

PACS: 12.39.Ki, 12.40.Ee, 12.40.Yx, 14.40.-n
\bigskip
\section{Introduction}
The spectrum of the scalar meson nonet is a long-standing problem of light 
meson spectroscopy. The number of resonances found in 
the region of 1--2 GeV exceeds the number of states that conventional quark 
model can accommodate \cite{pdg}. Extra states are interpreted alternatively 
as $K\bar{K}$ molecules, glueballs, multi-quark states or hybrids. In 
particular, except for a well established scalar isodoublet state, the 
$K_0^\ast (1430),$ the Particle Data Group (PDG) \cite{pdg} lists two 
isovector states, the $a_0(980)$ and $a_0(1450).$ The latter, having mass and 
width 1450$\pm 40$ MeV, $270\pm 40$ MeV, respectively, was discovered recently
by the Crystal Barrel collaboration at LEAR \cite{LEAR}. The third isovector
state (not included in \cite{pdg}), $a_0(1320),$ having mass and width $1322
\pm 30$ MeV, $130\pm 30$ MeV, respectively, was 
seen by GAMS \cite{GAMS}. There are four isoscalar states in \cite{pdg}, the 
$f_0(400-1200)$ (or $\sigma ),$ the interpretation of which as a particle is 
controversial due to a huge width of 600--1000 MeV, $f_0(980),$ $f_0(1370)$ 
(which stands for two separate states, $f_0(1300)$ and $f_0(1370)$, of a 
previous edition of PDG \cite{pdg1}), and $f_0(1500)$ (which also stands for
two separate states, $f_0(1525)$ and $f_0(1590),$ of a previous edition of
PDG), and two more possibly scalar states, the $f_J(1710),$ $J=0$ or 2, seen 
in radiative $J/\Psi $ decays, and an $\eta $-$\eta $ resonance $X(1740)$ 
whose spin is also uncertain, produced in $p\bar{p}$ annihilation in flight 
and in charge-exchange. Recently several groups claimed different scalar 
isoscalar structures close to 1500 MeV, including a narrow state with mass
$1450\pm 5$ MeV and width $54\pm 7$ MeV \cite{Abatzis}. The lightest of the 
three states at 1505 MeV, 1750 MeV and 2104 MeV revealed upon reanalyzing of 
data on $J/\Psi \rightarrow \gamma 2\pi ^{+}2\pi ^{-}$ \cite{Bugg}, and the 
$f_0(1450),$ $f_0(1500),$ $f_0(1520).$ The masses, widths and decay branching 
ratios of these states are incompatible within the errors quoted by the groups.
We do not consider it as plausible that so many scalar isoscalar states exist 
in such a narrow mass interval. Instead, we take the various states as 
manifestation of one object which should be identified with the $f_0(1500)$ of
recent PDG (which, however, may have a higher mass of about 1550 MeV).

It has been convincingly argued that the narrow $a_0(980)$, which has also 
been seen as a narrow structure in $\eta \pi $ scattering, can be generated by
meson-meson dynamics alone \cite{KK}. This interpretation of the $a_0(980)$ 
leaves the $a_0(1320)$ or $a_0(1450)$ (which may be manifestations of one 
state having a mass in the interval 1350-1400 MeV) as the 1 $^3P_0$ $q\bar{q}$
state. Similarly, it is mostly assumed that the $f_0(980)$ is a $K\bar{K}$ 
molecule. The mass degeneracy and their proximity to the $K\bar{K}$ threshold 
seem to require that the nature of both, $a_0(980)$ and $f_0(980),$ states 
should be the same. On the other hand, the $K\bar{K}$ interaction in the $I=1$
and $I=0$ channels is very different: the extremely attractive $I=0$ 
interaction may not support a loosely bound state. Instead, it may just 
define the pole position of the $f_0(980)$ $q\bar{q}$ resonance. Indeed, 
Morgan and Pennington \cite{MP} find the $f_0(980)$ pole structure 
characteristic for a genuine resonance of the constituents and not of a weakly
bound system. The $I=1$ $K\bar{K}$ interaction is weak and may generate a 
$K\bar{K}$ molecule. Alternatively, T\"{o}rnqvist \cite{Torn}
interprets both the $f_0(980)$ and $a_0(980)$ as the members of the $q\bar{q}$
nonet with strong coupling to the decay channels, which, however, does not 
account for the recently discovered $a_0(1320)$ and $a_0(1450).$

With respect to the $f_0(1370)$ (or two separate states, $f_0(1300)$ and $f_0(
1370),$ according to a previous edition of PDG), we follow the arguments of 
Morgan and Pennington \cite{MP} and assume that the $\pi \pi $ interaction 
produces both very broad, $f_0(1000),$ and narrow, $f_0(980),$ states, giving 
rise to a dip at 980 MeV in the squared $\pi \pi $ scattering amplitude $T_{11
}.$ In this picture, the $f_0(1370)$ is interpreted as the high-mass part of 
the $f_0(1000)$ (the low-mass part may be associated with the $\sigma $ of 
recent PDG). In experiments, the $f_0(1000)$ shows up at $\sim 1300$ MeV 
because of the pronounced dip in $|T_{11}|^2$ at $\sim 1$ GeV. The $f_0(1000)$
has an extremely large width; thus the resonance interpretation is 
questionable. It could be generated by $t$-channel exchanges instead of 
inter-quark forces \cite{ZB}.

The $f_0(1500)$ state has a peculiar decay pattern \cite{AC} 
\beq
\pi \pi :\eta \eta :\eta \eta ^{'}:K\bar{K}=1.45:0.39\pm 0.15:0.28\pm 0.12:
<0.15. 
\eeq
This pattern can be reproduced by assuming the existence of a further scalar 
state which is mainly $s\bar{s}$ and should have a mass of about 1700 MeV, 
possibly the $f_J(1710),$ and tuning the mixing of the $f_0(1500)$ with the 
$f_0(1370)$ $n\bar{n}$ and the (predicted) $f_0(1700)$ $s\bar{s}$ states 
\cite{AC}. In this picture, the $f_0(1500)$ is interpreted as a glueball state
with strong mixing with the close-by conventional scalar mesons. The 
interpretation of the $f_0(1500)$ as a conventional $q\bar{q}$ state, as well 
as the qualitative explanation of its reduced $K\bar{K}$ partial width, were 
given by Klempt {\it et al.} \cite{Klempt} in a relativistic quark model with 
linear confinement and an instanton-induced interaction. The quantitative 
explanation of the reduced $K\bar{K}$ partial width of the $f_0(1500)$ was 
given in a very recent publication by the same authors \cite{Klempt1}.

The above arguments lead one to the following spectrum of the scalar meson 
nonet:
\bqry
a_0(1320)\;\;{\rm or}\;\;a_0(1450), & K_0^\ast (1430), \NL
f_0(980)\;\;{\rm or}\;\;f_0(1000), & f_0(1500).
\eqry
This spectrum agrees essentially with the $q\bar{q}$ assignments found by 
Klempt {\it et al.} \cite{Klempt}, and Dmitrasinovic \cite{Dmitra} who
considered the Nambu--Jona-Lasinio model with a $U_A(1)$ breaking 
instanton-induced 't Hooft interaction. The spectrum of the 
meson nonet given in \cite{Klempt} is
\beq
a_0(1320),\;K_0^\ast (1430),\;f_0(1470),\;f_0^{'}(980),
\eeq
while that suggested by Dmitrasinovic, on the basis of the sum rule
\beq
m_{f_0}^2+m_{f_0^{'}}^2+m_\eta ^2+m_{\eta ^{'}}^2=2(m_K^2+m_{K_0^\ast }^2)
\eeq
derived in his paper, is \cite{Dmitra}
\beq
a_0(1320),\;K_0^\ast (1430),\;f_0(1590),\;f_0^{'}(1000).
\eeq
In this paper we show that similar results are reproduced by the application 
of the linear mass spectrum of a multiplet, discussed in a series of papers by
Larry Horwitz and his collaborators \cite{series,spectrum}, to a composite 
system of the pseudoscalar and scalar meson nonets. The three mass relations 
that we obtain (one of these is the Dmitrasinovic sum rule (4), another its 
two-flavor version) enables one to calculate the unknown masses of the $a_0,$ 
$f_0$ and $f_0^{'}$ states explicitly, in terms of the known masses of the 
pseudoscalar and scalar isodoublet states. The results turn out to agree 
essentially with the spectrum (2).

\section{Hadronic resonance spectrum}
The hadronic mass spectrum is an essential ingredient in theoretical 
investigations of the physics of strong interactions. It is well known that 
the correct thermodynamic description of hot hadronic matter requires 
consideration of higher mass excited states, the resonances, whose 
contribution becomes essential at temperatures $\sim O(100$ MeV) 
\cite{Shu,Leut}. The method for taking into account these 
resonances was suggested by Belenky and Landau \cite{BL} as considering 
unstable particles on an equal footing with the stable ones in the 
thermodynamic quantities; e.g., the formulas for the pressure and energy 
density in a resonance gas read\footnote{For simplicity, we neglect the 
chemical potential and approximate the particle statistics by the 
Maxwell-Boltzmann one.}
\beq
p=\sum _ip_i=\sum _ig_i\frac{m_i^2T^2}{2\pi ^2}K_2\left(\frac{m_i}{T}\right),
\eeq
\beq
\rho =\sum _i\rho _i,\;\;\;\rho _i=T\frac{dp_i}{dT}-p_i,
\eeq
where $g_i$ are the corresponding degeneracies ($J$ and $I$ are spin and
isospin, respectively), $$g_i=\frac{\pi ^4}{90}\times \left[
\begin{array}{ll}
(2J_i+1)(2I_i+1) & {\rm for\;non-strange\;mesons} \\
4(2J_i+1) & {\rm for\;strange}\;(K)\;{\rm mesons} \\
2(2J_i+1)(2I_i+1)\times 7/8 & {\rm for\;baryons}
\end{array} \right. $$
These expressions may be rewritten with the help of a {\it resonance spectrum,}
\beq
p=\int _{m_1}^{m_2}dm\;\tau (m)p(m),\;\;\;p(m)\equiv \frac{m^2T^2}{2\pi ^2}
K_2\left(\frac{m}{T}\right),
\eeq
\beq
\rho =\int _{m_1}^{m_2}dm\;\tau (m)\rho (m),\;\;\;\rho (m)\equiv 
T\frac{dp(m)}{dT}-p(m),
\eeq
normalized as 
\beq
\int _{m_1}^{m_2}dm\;\tau (m)=\sum _ig_i,
\eeq
where $m_1$ and $m_2$ are the masses of the lightest and heaviest species, 
respectively, entering the formulas (6),(7). 

In both the statistical bootstrap model \cite{Hag,Fra} and the dual resonance 
model \cite{FV}, a resonance spectrum takes on the form
\beq
\tau (m)\sim m^a\;e^{m/T_0},
\eeq
where $a$ and $T_0$ are constants. The treatment of a hadronic resonance gas 
by means of the spectrum (11) leads to a singularity in the thermodynamic 
functions at $T=T_0$ \cite{Hag,Fra} and, in particular, to an infinite number
of the effective degrees of freedom in the hadron phase, thus hindering a
transition to the quark-gluon phase. Moreover, as shown by Fowler and Weiner
\cite{FW}, an exponential mass spectrum of the form (11) is incompatible with
the existence of the quark-gluon phase: in order that a phase transition from 
the hadron phase to the quark-gluon phase be possible, the hadronic spectrum
cannot grow with $m$ faster than a power. 

In ref. \cite{spectrum} a model for a transition from a phase of strongly 
interacting hadron constituents to the hadron phase described by a resonance 
spectrum, Eqs. (8),(9), was considered. The strongly interacting phase of 
hadron constituents was described by a manifestly covariant relativistic 
statistical mechanics which has been developed by Larry Horwitz and his 
collaborators and turned out to be a reliable framework in the description of 
realistic physical systems like strongly interacting matter \cite{mancov}.
An example of such a transition may be a relativistic high temperature
Bose-Einstein condensation studied by the same authors in ref. \cite{cond}, 
which corresponds, in the way suggested by Haber and Weldon \cite{HW}, to 
spontaneous flavor symmetry breakdown, $SU(3)_F\rightarrow SU(2)_I\times 
U(1)_Y,$ upon which hadronic multiplets are formed, with the masses obeying 
the Gell-Mann--Okubo formulas \cite{GMO}
\beq
m^\ell =a+bY+c\left[ \frac{Y^2}{4}-I(I+1)\right];
\eeq
here $I$ and $Y$ are the isospin and hypercharge, respectively, $\ell $ is 2 
for mesons and 1 for baryons, and $a,b,c$ are independent of $I$ and $Y$ but, 
in general, depend on $(p,q),$ where $(p,q)$ is any irreducible representation 
of $SU(3).$ Then only the assumption on the overall degeneracy being conserved 
during the transition is required to lead to the unique form of a resonance 
spectrum in the hadron phase:
\beq
\tau (m)=Cm,\;\;\;C={\rm const}.
\eeq
Zhirov and Shuryak \cite{ZS} have found the same result on phenomenological 
grounds. As shown in ref. \cite{ZS}, the spectrum (13), used in the  
formulas (8),(9) (with the upper limit of integration infinity), leads to
the equation of state $p,\rho \sim T^6,$ $p=\rho /5,$ called by Shuryak the 
``realistic'' equation of state for hot hadronic matter \cite{Shu}, which has
some experimental support. Zhirov and Shuryak \cite{ZS} have calculated 
the velocity of sound, $c_s^2\equiv dp/d\rho =c_s^2(T),$ with $p$ and $\rho $ 
defined in Eqs. (6),(7), and found that $c_s^2(T)$ at first increases with $T$
very quickly and then saturates at the value of $c_s^2\simeq 1/3$ if only the
pions are taken into account, and at $c_s^2\simeq 1/5$ if resonances up to 
$M\sim 1.7$ GeV are included. 

The agreement of the results given by the linear spectrum (13)
with those obtained directly from Eq. (6) for the actual hadronic species with
the corresponding degeneracies, for all well-established multiplets, both 
mesonic and baryonic, was checked in ref. \cite{series} and found to be
excellent. Thus, the theoretical implication that a linear spectrum is the 
actual spectrum in the description of individual hadronic multiplets, is 
consistent with experiment as well. 

The easiest way to see that a linear spectrum corresponds to the actual 
spectrum of a meson nonet is as follows. The average mass squared for a 
spin-$s$ nonet is
\beq
\langle m^2\rangle \equiv \frac{\sum _ig_i\;m_i^2}{\sum_ig_i}=\frac{3m_1^2+
4m_{1/2}^2+m_8^2+m_9^2}{9},
\eeq
where $m_1,\;m_{1/2},\;m_8,\;m_9$ are the masses of isovector, isodoublet,
isoscalar octet and isoscalar singlet states, respectively, and the spin 
degeneracy, $2s+1,$ cancels out. Since $m_9^2=\langle m^2\rangle $ 
\cite{series}, Eq. (14) may be rewritten as  
\beq
\langle m^2\rangle =\frac{3m_1^2+4m_{1/2}^2+m_8^2}{8},
\eeq
which is the average mass squared of the octet. The same average mass squared
can be also calculated with the help of the linear spectrum, as 
\beq
\langle m^2\rangle =\frac{\int _{m_1}^{m_8}dm\;m^3}{\int _{m_1}^{m_8}dm\;m}=
\frac{(m_8^4-m_1^4)/4}{(m_8^2-m_1^2)/2}=\frac{m_1^2+m_8^2}{2}.
\eeq
By equating (15) and (16), one obtains
\beq
4m_{1/2}^2=3m_8^2+m_1^2,
\eeq
which is the Gell-Mann--Okubo mass formula for an octet (as follows from (12)).
In general, the isoscalar octet and singlet states get mixed, because of 
$SU(3)$ breaking, which results in the physical $\omega _{0^{'}}$ and $\omega 
_{0^{''}}$ states (the $\omega _{0^{'}}$ is a mostly octet isoscalar): $$\omega
_{0^{'}}=\omega _8\cos \theta _M-\omega _9\sin \theta _M,$$ $$\omega _{0^{''}}
=\omega _8\sin \theta _M+\omega _9\cos \theta _M,$$ where $\theta _M$ is a 
mixing angle. Assuming that the matrix element of the Hamiltonian between the 
states yields a mass squared, i.e., $m_{0^{'}}^2=\langle \omega _{0^{'}}|H|
\omega _{0^{'}}\rangle $ etc., one obtains from the above relations \cite{Per},
\beq
m_{0^{'}}^2=m_8^2\cos ^2\theta _M+m_9^2\sin ^2\theta _M-2m_{89}^2\sin \theta _M
\cos \theta _M,
\eeq 
\beq
m_{0^{''}}^2=m_8^2\sin ^2\theta _M+m_9^2\cos ^2\theta _M+2m_{89}^2\sin \theta
_M\cos \theta _M.
\eeq
Since $\omega _{0^{'}}$ and $\omega _{0^{''}}$ are orthogonal, one has further
\beq
m_{0^{'}0^{''}}^2=0=(m_8^2-m_9^2)\sin \theta _M\cos \theta _M+m_{89}^2(\cos ^2
\theta _M-\sin ^2\theta _M).
\eeq
Eliminating $m_9$ and $m_{89}$ from (18)-(20) yields
\beq
\tan ^2\theta _M=\frac{m_8^2-m_{0^{'}}^2}{m_{0^{''}}^2-m_8^2}.
\eeq
It follows from (18),(19) that, independent of $\theta _M,$ $m_{0^{'}}^2+
m_{0^{''}}^2=m_8^2+m_9^2,$ and therefore, as seen in Eq. (14), the average 
mass squared of the nonet does not change under the mixing of the $\omega _8$
and $\omega _9.$ This fact is easily understood in a manifestly covariant 
theory in which a total mass squared is rigorously conserved \cite{mancov}: an
average mass squared is equal to a total mass squared devided by the total 
number of degrees of freedom (e.g., 9 for a meson nonet); the conservation of
the latter will mean that an average mass squared is conserved as well. 
 
As also follows from (18),(19), $m_8^2=m_{0^{'}}^2 \cos ^2 \theta _M+m_{0^{
''}}^2 \sin ^2\theta _M,$ and therefore, Eq. (17) may be rewritten as
\beq
4m_{1/2}^2-m_1^2-3m_{0^{'}}^2=3\left( m_{0^{''}}^2-m_{0^{'}}^2\right) \sin ^2
\theta _M,
\eeq
which is the Sakurai mass formula \cite{Sakurai}. For an octet-singlet mixing
close to ``ideal'' one, for which $\omega _{0^{'}}\simeq s\bar{s},$ $\omega
_{0^{''}}\simeq (u\bar{u}+d\bar{d})/\sqrt{2},$ $\sin \theta _M\simeq \sqrt{
1/3};$ it then follows from (22) that
\beq
4m_{1/2}^2\cong m_1^2+m_{0^{''}}^2+2m_{0^{'}}^2,
\eeq
which is a formula for the ``ideal'' structure of a nonet. As discussed in 
\cite{series}, this is the only mass relation for a mixed nonet consistent with
the linear spectrum and the conservation of the average (and total) mass 
squared. This relation also appears in the algebraic approach to QCD \cite{OT}
as the only solution to an infinite chain of constraints on the masses of the
nonet states imposed by the charge algebras and the requirement of the 
asymptotic $SU(3)$ flavor symmetry. One sees that the linear spectrum requires
that the masses of the members of a given meson nonet satisfy the formula (23).
We have found that Eq. (23) agrees with the experimentally established meson
masses, with an accuracy of up to 3\%, for all well-established meson nonets,
except for the pseudoscalar and, possibly, scalar ones. As is generally 
believed, the reason for the invalidity of Eq. (23) for the pseudoscalar nonet
is a large dynamical mass of the $\eta _9$ due to axial $U(1)$ symmetry
breakdown developed before it mixes with the $\eta _8$ to form the physical
$\eta $ and $\eta ^{'}$ states. The well known solution to the $U_A(1)$ problem
suggested by 't Hooft \cite{tH} is based on instanton effects. 't Hooft has
shown that an expansion of the (euclidian) action around the one-instanton 
solutions of the gauge fields assuming dominance of the zero modes of the 
fermion fields leads to an effective $2N_f$-fermion interaction $(N_f$ being 
the number of fermion flavors) not covered by perturbative gluon exchange, 
which gives additional contribution to ordinary confinement quark-antiquark 
interaction. As shown in ref. \cite{Munz}, due to its point-like nature and 
specific spin structure, the instanton-induced interaction in the formulation 
of 't Hooft acts on the states with spin zero only. The masses of the other 
mesons with non-vanishing spin are therefore determined by the confinement 
interaction alone leading to the conventional splitting and an ideal mixing of
the $q\bar{q}$ nonets which results in the mass spectra (23). The only two 
nonets whose mass spectra turn out to be affected by an instanton-induced 
interaction are spin zero pseudoscalar and scalar nonets. Quantitatively, an 
instanton-induced interaction for the scalar mesons is of the same magnitude
but opposite in sign, in comparison with the pseudoscalars \cite{Klempt}. 
It, therefore, lowers the masses of the scalar states with a dominantly flavor 
singlet structure, and rises the masses of the scalar states with a dominantly 
flavor octet structure, just in contrast to the case of the pseudoscalar 
isoscalar flavor singlet and octet states whose masses are pushed up and down,
respectively. Thus, the physical mass spectra of the pseudoscalar and scalar 
meson nonets are generated by the following mechanism provided by an
instanton-induced interaction (in what follows, $\pi $ stands for the mass of 
the pion, etc.):

for the pseudoscalar nonet:

$$(\pi ,\;K,\;\eta _8,\;\eta _9)\;\longrightarrow \;(\pi ,\;K,\;\eta _8^{'}
\leq \eta _8,\;\eta _9^{'}\geq \eta _9)$$ $$\longrightarrow ({\rm mixing\;\;of}
\;\;\eta _8^{'}\;\;{\rm and}\;\;\eta _9^{'})\longrightarrow (\pi ,\;K,\;\eta ,
\;\eta ^{'});$$

for the scalar nonet: 
    
$$(a_0,\;K_0^\ast ,\;f_8,\;f_9)\;\longrightarrow \;(a_0,\;K_0^\ast ,\;f_8^{'}
\geq f_8,\;f_9^{'}\leq f_9)$$ $$\longrightarrow ({\rm mixing\;\;of}\;\;f_8^{'}
\;\;{\rm and}\;\;f_9^{'})\longrightarrow (a_0,\;K_0^\ast ,\;f_0,\;f_0^{'}).$$
The parameters of the 't Hooft interaction may be fixed to reproduce the 
physical spectrum of the pseudoscalar states \cite{Blask}; subsequent 
application to the scalar states yields the spectrum (3) \cite{Klempt}. In this
paper we consider the spectra of the pseudoscalar and scalar states in a
manifestly covariant framework, by the application of the linear mass spectrum
to the composite system of both the pseudoscalar and scalar meson nonets.

\section{Relations for the masses of the pseudoscalar and scalar states}
We adopt the following pattern: both the pseudoscalar and scalar mesons emerge
upon nucleation from the strongly interacting ensemble of hadron constituents
to quasi-discrete (hadronic) mass levels, as the unique composite system which
exhibits intermultiplet (instanton-induced) interaction. Let us denote the 
total shifts of the squared masses of the isoscalar states of the pseudoscalar
and scalar nonets (which are now the subsystems of our composite system) which
arise as a result of this interaction, as discussed above, as
$\triangle _{PS}$ and $\triangle _S,$ respectively:
\beq
\triangle _{PS}\equiv \eta _8^{'2}+\eta _9^{'2}-\eta _8^2-\eta _9^2,
\eeq 
\beq
\triangle _S\equiv f_8^{'2}+f_9^{'2}-f_8^2-f_9^2.
\eeq   
Since, independent of the mixing angle of the corresponding mixings, 
\beq
\eta _8^{'2}+\eta _9^{'2}=\eta ^2+\eta ^{'2},
\eeq
\beq
f_8^{'2}+f_9^{'2}=f_0^2+f_0^{'2},
\eeq
Eqs. (24) and (25) may be rewritten as
\beq
\triangle _{PS}\equiv \eta ^2+\eta ^{'2}-\eta _8^2-\eta _9^2,
\eeq 
\beq
\triangle _S\equiv f_0^2+f_0^{'2}-f_8^2-f_9^2.
\eeq
It can be easily shown that $\triangle _{PS}$ and $\triangle _S$ are of the 
same magnitude but opposite in sign,
\beq
\triangle _{PS}=-\triangle _S,
\eeq
just as the instanton-induced interaction for the scalar mesons as compared to
that for the pseudoscalars, as discussed above. Indeed, independent of the 
details of the intermultiplet interaction (which is responsible only for the
physical mass spectrum of a multiplet), the total mass squared of the composite
system we are considering is rigorously conserved. Therefore, by equating the
total mass squared of the bare (noninteracting) system and that of the physical
pseudoscalar and scalar nonets, one obtains (taking into account the 
corresponding degeneracies) $$3\pi ^2+4K^2+\eta _8^2+\eta _9^2
+3a_0^2+4K_0^{\ast 2}+f_8^2+f_9^2=3\pi ^2+4K^2+\eta ^2+\eta ^{'2}+3a_0^2+4K_0^{
\ast 2}+f_0^2+f_0^{'2},$$ or $$\eta ^2+\eta ^{'2}-\eta _8^2-\eta _9^2=f_8^2+
f_9^2-f_0^2-f_0^{'2},$$ which reduces to Eq. (30), in view of (26)-(29). As 
discussed in \cite{series} on the basis of the linear spectrum of a meson 
nonet, the squared mass of the isoscalar singlet state of a nonet is given by 
(16), $$m_9^2=\langle m^2\rangle =\frac{m_1^2+m_8^2}{2},$$
which reduces, through (17), to \cite{series} $$m_8^2+m_9^2=2m_{1/2}^2.$$
In the case we are considering here, for the bare system of the two nonets one 
has, therefore, the two relations,
\beq
\eta _8^2+\eta _9^2=2K^2,
\eeq
\beq
f_8^2+f_9^2=2K_0^{\ast 2},
\eeq
which for the interacting system transform into 
\beq
\eta ^2+\eta ^{'2}=2K^2+\triangle _{PS},
\eeq
\beq
f_0^2+f_0^{'2}=2K_0^{\ast 2}+\triangle _S,
\eeq
in view of (24)-(27). Now, summing up (33) and (34) and using (30), one obtains
\beq
\eta ^2+\eta ^{'2}+f_0^2+f_0^{'2}=2\left( K^2+K_0^{\ast 2}\right) ,
\eeq
which is the Dmitrasinovich sum rule (4) and which represents the first of the
three mass relations we need in order to determine the unknown masses
of the scalar mesons. The remaining two relations are obtained by the 
application of the linear spectrum to the bare system and that of the two 
physical nonets, respectively.

For the bare system of the two nonets which occupies the mass interval $(\pi ,
f_8),$ by representing the average mass squared in two different ways, as was
done in a previous section for the derivation of the Gell-Mann--Okubo mass 
formula (17),
\beq
\frac{3\pi ^2+4K^2+\eta _8^2+\eta _9^2+3a_0^2+4K_0^{\ast 2}+f_8^2+f_9^2}{18}=
\frac{\pi ^2+f_8^2}{2},
\eeq
and using Eqs. (31),(32) for both, $\eta _8^2+\eta _9^2$ and $f_8^2+f_9^2,$ in
the l.h.s. of (36), and the Gell-Mann--Okubo formula
\beq
f_8^2=\frac{4K_0^{\ast 2}-a_0^2}{3}
\eeq
in the r.h.s. of (36), one finally arrives at
\beq
K_0^{\ast 2}-a_0^2=K^2-\pi ^2,
\eeq
which is the second mass relations we need. A relation of this type may be
anticipated on the basis of the relations $$K^{\ast 2}-\rho ^2=K^2-\pi ^2,\;\;
\;K_2^{\ast 2}-a_2^2=K^2-\pi ^2,\;\;\;{\rm etc.,}$$ provided by either the 
algebraic approach to QCD developed in ref. \cite{OT} or phenomenological 
formulas $$m_1^2=2Bm_n+C,\;\;\;m_{1/2}^2=B(m_n+m_s)+C$$ (where $B$ is related 
to the quark condensate, $C$ is a  constant within a given meson nonet, and 
$m_n$ and $m_s$ are the masses of non-strange and strange quarks, 
respectively) motivated by the linear mass spectrum of a nonet and the 
collinearity of Regge trajectories of the corresponding $I=1$ and $I=1/2$ 
states, as discussed in refs. \cite{series,NPA}, which reduce to $$m_{1/2}^2-
m_1^2=B(m_s-m_n),$$ independent of the quantum numbers of a nonet (for lower
spin nonets, at least). 
 
To obtain the third mass relation, we consider the system of the two physical
nonets. As discussed above, in the case of a confinement interaction alone 
leading to an ideal mixture of a $q\bar{q}$ nonet, the physical mass spectrum
of a nonet takes the form (23) which is, on the other hand, the manifestation
of the conservation of the average (and total, in this case) mass squared. In 
the case we are considering here, the average squared masses of each of the 
two nonets are obviously not conserved under the mass shifts caused by an 
instanton-induced interaction, but the average mass squared of the composite 
system is conserved, due to the conservation of both the total mass squared and
the total number of degrees of freedom. We, therefore, assume the validity of
the two relations,
\beq
4K^2=\pi ^2+\eta ^{'2}+2\eta ^2-\triangle ^{'2},
\eeq
\beq
4K_0^{\ast 2}=a_0^2+f_0^{'2}+2f_0^2+\triangle ^{'2},
\eeq
which, upon summing up, lead to the formula
\beq
4\left( K^2+K_0^{\ast 2}\right) =\pi ^2+a_0^2+\eta ^{'2}+f_0^{'2}+2\left( \eta
^2+f_0^2\right) ,
\eeq
consistent with the linear spectrum of the composite system of the two nonets 
and the conservation of the average mass squared of this composite system, 
just as Eq. (23) is consistent with the linear mass spectrum and the conserved 
average mass squared of an individual ideally mixed nonet. It then follows 
from (35) and (41) that
\beq
\eta ^{'2}-\pi ^2=a_0^2-f_0^{'2},
\eeq
which is the third relation we need in order to establish the unknown masses of
the scalar mesons. The above relation was obtained by Dmitrasinovic in the 
two-flavor version of the Nambu--Jona-Lasinio model with the instanton-induced
't Hooft interaction \cite{Dmitra}. It may be also derived by using the linear
mass spectrum in the two-flavor case. Indeed, in this case a $q\bar{q}$ meson 
multiplet is a quadruplet composed with three isovector and one isoscalar 
states which are mass degenerate in the ``ideal'' case of the confinement 
interaction alone, since they are composed with the same non-strange quarks 
(also, the standard procedure of representing the average mass squared in two 
different ways which we are using in this paper gives now $(3m_1^2+m_0^2)/4=
(m_1^2+m_0^2)/2,$ which reduces to $m_1=m_0).$ In the case of the 
instanton-induced interaction which perturbs the ideal structure of a
multiplet, one has the physical mass spectrum of the composite system of the 
two pseudoscalar and scalar quadruplets, $\pi, \eta^{'},a_0,f_0^{'}$ $(\eta ^{
'}\geq \pi ,\;f_0^{'}\leq a_0).$ Now, the procedure of representing the average
mass squared in two different ways gives $$\frac{3\pi ^2+\eta ^{'2}+3a_0^2+f_
0^{'2}}{8}=\frac{\pi ^2+a_0^2}{2},$$ which reduces to Eq. (42). 

The three relations, (35),(38) and (42), obtained above enables one to 
determine the unknown masses of the scalar mesons explicitely, in terms of the
known masses of the pseudoscalar and $K_0^\ast $ mesons:
\bqry
a_0^2 & = & K_0^{\ast 2}-K^2+\pi ^2, \\
f_0^2 & = & K_0^{\ast 2}+3K^2-2\pi ^2-\eta ^2, \\
f_0^{'2} & = & K_0^{\ast 2}-K^2+2\pi ^2-\eta ^{'2}.
\eqry
With the physical masses of these states (in MeV) \cite{pdg}, $\pi =138,$ $K=
495,$ $\eta =547,$ $\eta ^{'}=958,$ $K_0^\ast =1429,$ the formulas (43)-(45) 
give
\bqry      
a_0 & = & 1348, \\
f_0 & = & 1562, \\
f_0^{'} & = & 958.
\eqry
One sees that the obtained values for the masses of the scalar mesons agree
essentially with the spectrum (2). The predicted mass of the isoscalar mostly 
singlet state (958) is within 2\% of the physical mass of the $f_0(980)$ meson
which is $980\pm 10$ MeV \cite{pdg}, and is not too far from the mass of the 
$f_0(1000)$ meson. On the basis of the results obtained in this paper, we 
tentatively adopt the following $q\bar{q}$ assignment for the scalar meson 
nonet: $$a_0(1320),\;\;\;K_0^\ast (1430),\;\;\;f_0(1550),\;\;\;f_0^{'}(980).$$
In order to obtain the mixing angle which corresponds to this assignment, we 
first calculate the mass of the $f_8$ using the Gell-Mann--Okubo mass formula
and Eq. (43):
\beq
f_8^2=\frac{4K_0^{\ast ^2}-a_0^2}{3}=K_0^{\ast 2}+\frac{K^2-\pi ^2}{3},
\eeq
which gives $$f_8=1455.$$ Assuming now that the shift of the masses of the 
isoscalar octet states caused by an instanton-induced interaction is small
compared to that of the masses of the isoscalar singlet states, i.e., $f_8^{'}
\simeq f_8,$ one has, with the help of (21),
\beq
\tan ^2\theta _M\simeq \frac{f_8^2-f_0^2}{f_0^{'2}-f_8^2},
\eeq
which gives $$\theta _M\simeq 26.5^{o}.$$ This value is in good quantitative 
agreement with that predicted by Ritter {\it et al.} \cite{Klempt1}, $\theta _M
\approx 25^{o},$ for which the partial widths of the $f_0(1500)$ calculated in
their paper are $$\pi \pi :\eta \eta :\eta \eta ^{'}:K\bar{K}=1.45:0.32:0.18:
0.03,$$ in excellent agreement with the experimentally observed partial widths,
Eq. (1).  

\section{Concluding remarks}
We have obtained three mass relations for the scalar states, by the application
of the linear mass spectrum to a composite system of both the pseudoscalar and
scalar meson nonets, which enable one to calculate the masses of the $a_0,$ 
$f_0$ and $f_0^{'}$ mesons explicitly, in terms of the known masses of the
pseudoscalar and $K_0^\ast $ mesons, and arrive at the $q\bar{q}$ assignment
for the scalar meson nonet (2). 

A question remains about which of the two, $f_0(980)$ or $f_0(1000),$ states
should be associated with the predicted isoscalar mostly singlet state having 
a mass in the vicinity of 1 GeV. The corresponding decision is difficult to
make on the basis of a naive quark model. However, two observations support the
interpretation of the $f_0(980)$ as a $q\bar{q}$ state. First, the 
$t$-dependence of the $f_0(980)$ and the broad background produced in $\pi ^{-}
p\rightarrow \pi ^0\pi ^0n$ differ substantially \cite{Alde}. The $f_0(1000)$
is produced in peripheral collisions only, while the $f_0(980)$ shows a 
$t$-dependence, as expected for a $q\bar{q}$ state. Second, since the scalar
nonet is not ideally mixed, the both, $n\bar{n}$ and $s\bar{s},$ components of 
the isoscalar mostly singlet state should be appreciable. The $f_0(980)$ is 
seen strongly in $J/\Psi \rightarrow \phi f_0(980),$ but at most weakly in 
$J/\Psi \rightarrow \omega f_0(980).$ On the basis of quark diagrams, one must
conclude that the $f_0(980)$ has a large $s\bar{s}$ component; its decay into 
$\pi \pi $ with the corresponding branching ratio 78\% \cite{pdg} underlines an
appreciable $n\bar{n}$ component.

Although our results need experimental confirmation, the fact that the three 
different approaches developed in refs. \cite{Klempt,Dmitra} and the present
paper lead to essentially the same $q\bar{q}$ assignment for the scalar meson
nonet should imply that we are not too far from the physics of a real world.

\section*{Acknowledgements}
The author wishes to thank T. Goldman for very valuable discussions during the
preparation of this paper.

\end{document}